\documentclass[%
  prb,
  aps,
  twocolumn,
  superscriptaddress,
  showpacs
]{revtex4-1}

\usepackage[%
  colorlinks=true,
  linkcolor=black, 
  citecolor=blue, 
  urlcolor=blue, 
  unicode=true
]{hyperref}

\usepackage{graphicx,amssymb}

\begin{document}

%\preprint{}


\title{Metastable and localized Ising magnetism in $\alpha$-CoV$_{2}$O$_{6}$ magnetization plateaus}

\author{L. Edwards}
\affiliation{School of Physics and Astronomy, Cardiff University, Cardiff CF24 3AA, UK}
\affiliation{ISIS Facility, Rutherford Appleton Laboratory, Chilton, Didcot OX11 0QX, UK}
\author{H. Lane}
\affiliation{ISIS Facility, Rutherford Appleton Laboratory, Chilton, Didcot OX11 0QX, UK}
\affiliation{School of Chemistry and Centre for Science at Extreme Conditions, University of Edinburgh, Edinburgh EH9 3FJ, UK}
\affiliation{School of Physics and Astronomy, University of Edinburgh, Edinburgh EH9 3JZ, UK}
\author{A. M. Arevalo-Lopez}
\affiliation{Univ. Lille, CNRS, Centrale Lille, Univ. Artois, UMR 8181 – UCCS – Unité de Catalyse et Chimie du Solide, F-59000 Lille, France.}
\author{M. Songvilay}
\affiliation{School of Physics and Astronomy, University of Edinburgh, Edinburgh EH9 3JZ, UK}
\author{E. Pachoud}
\affiliation{School of Chemistry, University of Edinburgh, Edinburgh EH9 3FJ, UK}
\author{Ch. Niedermayer}
\affiliation{Laboratory for Neutron Scattering, Paul Scherrer Institut, CH-5232 Villigen, Switzerland}
\author{G. Tucker}
\affiliation{Institute of Physics, \'Ecole Polytechnique F\'ed\'erale le de Lausanne (EPFL), CH-1015 Lausanne, Switzerland}
\affiliation{Laboratory for Neutron Scattering, Paul Scherrer Institut, CH-5232 Villigen, Switzerland}
\author{P. Manuel}
\affiliation{ISIS Facility, Rutherford Appleton Laboratory, Chilton, Didcot OX11 0QX, UK}
\author{C. Paulsen}
\affiliation{Institut N\'{e}el, C.N.R.S - Universit\'e Grenoble Alpes, BP 166, 38042 Grenoble, France}
\author{E. Lhotel}
\affiliation{Institut N\'{e}el, C.N.R.S - Universit\'e Grenoble Alpes, BP 166, 38042 Grenoble, France}
\author{J. P. Attfield}
\affiliation{School of Chemistry and Centre for Science at Extreme Conditions, University of Edinburgh, Edinburgh EH9 3FJ, UK}
\author{S. R. Giblin}
\affiliation{School of Physics and Astronomy, Cardiff University, Cardiff CF24 3AA, UK}
\author{C. Stock}
\affiliation{School of Physics and Astronomy, University of Edinburgh, Edinburgh EH9 3JZ, UK}

\date{\today}

\begin{abstract}

$\alpha$-CoV$_{2}$O$_{6}$ consists of $j_{\mathrm{eff}}={1 \over 2}$ Ising spins located on an anisotropic triangular motif with magnetization plateaus in an applied field.  We combine neutron diffraction with low temperature magnetization to investigate the magnetic periodicity in the vicinity of these plateaus.   We find these steps to be characterized by metastable and spatially short-range ($\xi\sim$ 10 \AA) magnetic correlations with antiphase boundaries defining a local periodicity of $\langle \hat{T}^{2} \rangle =\ \uparrow \downarrow$ to $\langle \hat{T}^{3} \rangle =\ \uparrow \uparrow \downarrow$, and $\langle \hat{T}^{4} \rangle=\ \uparrow \uparrow \downarrow \downarrow$ or $\uparrow \uparrow \uparrow \downarrow$ spin arrangements.  This shows the presence of spatially short range and metastable/hysteretic, commensurate magnetism in Ising magnetization steps. 

\end{abstract}

\pacs{}

\maketitle

\section{Introduction}

The magnetization of a classical Neel magnet in an applied field varies continuously with the spins ideally rotating with respect to the field direction to minimize energy.  In quantum antiferromagnets~\cite{Chubukov91:3,Zhit98:57} and in particular in the presence of frustrating magnetic interactions~\cite{Penc04:93}, the situation is more complex and steps in the magnetization can exist where the total $S^{z}_{total}=\sum_{i} S_{i}^{z}$ is unchanged in a varying magnetic field~\cite{Lacroix:book}.  We investigate the magnetization in a $j_{\mathrm{eff}}={1\over 2}$ magnet where a cascading series of hysteretic spatially short-range periodicities characterize the plateaus in the magnetization.

For insulating magnetic compounds, the presence of a magnetization plateau is indicative of an energetic gap.  This occurs in $S=1$ compounds where a Haldane gap is present~\cite{Haldane83:93A,Affleck87:59,Oshikawa00:84} in contrast to $S={1\over 2}$ systems where the excitation spectrum is gapless~\cite{Vasukuev18:3,Tennant93:70,Villain75:79}. While $S={1\over2}$ compounds are not expected to display magnetization plateaus based on energetics, dimerized $S={1\over 2}$ magnets do display a gap~\cite{Jacobs76:14,Hase93:70} and hence can show magnetization plateaus~\cite{Matsuda13:111}.  On more general grounds, based on the Lieb-Mattis-Schultz theorem~\cite{Lieb61:16}, it has been proposed that half-integer spin based compounds can display analogous Haldane gapped phases~\cite{Oshikawa97:78} and hence plateaus in the magnetization.  Based on this, spatially short ranged dilute-dimerized phases have been predicted in half-integer spins.~\cite{Totsuka98:57}  Other explanations for magnetization plateaus include the order by disorder~\cite{Henley89:62,Smirnov17:119,Agrestini11:106} mechanism where a ground state is selected through quantum fluctuations.  

Steps in the magnetization have been reported in many insulating and triangular based  compounds~\cite{Takagi95:64,Kikuchi05:94,Kageyama97:66,Kageyama97:66_2,Hardy03:15,Kudasov06:96,Sampath02:651,Niitaka01:87,Hardy06:74,Maignan04:14,Okamoto03:15,Rule08:100,Katsumata10:82,Ishii11:94,Ueda05:94} (and also see table in Ref. \onlinecite{Lacroix:book}) and also in materials with a continuous symmetry~\cite{Mitsuda00:69,Terada04:70,Fishman10:81,Fishmann11:106} where field induced magnetic transitions are observed.  In many of these examples, the magnetization steps are defined by phase transitions with spatially long-range correlations accompanied by new Bragg peaks, where neutron scattering data is available, and are understood by magnetic Hamiltonians with competing exchange interactions.  In this paper we discuss another situation where the magnetization plateaus are defined by short-range and metastable correlations.  

$\alpha$-CoV$_2$O$_6$ crystallizes in the centrosymmetric monoclinic space group $\mathrm{C}\, 2/m\, $, with lattice parameters ${a = 9.2283(1)}$, ${b = 3.50167(5)}$, ${c = 6.5983(1)\ \text{\AA}}$, ${\beta=112.0461(7)^{\circ}}$ differing from the triclinic $\gamma$-CoV$_{2}$O$_{6}$ polymorph.~\cite{Lenertz14:118}  The structure is based on Co$^{2+}$ ${(L=2,\, S={3 \over 2})}$ ions sited on an anisotropic triangular motif.~\cite{Fazekas18:9,Stock09:103,Giot07:99,Dally18:98,Dally18:9} The small exchange coupling between the spins relative to the anisotropy energy,~\cite{Wallington15:92} is suggestive of an underlying Ising/uniaxial symmetry. 

\begin{figure}[tb]
\centering
\includegraphics[width=75mm]{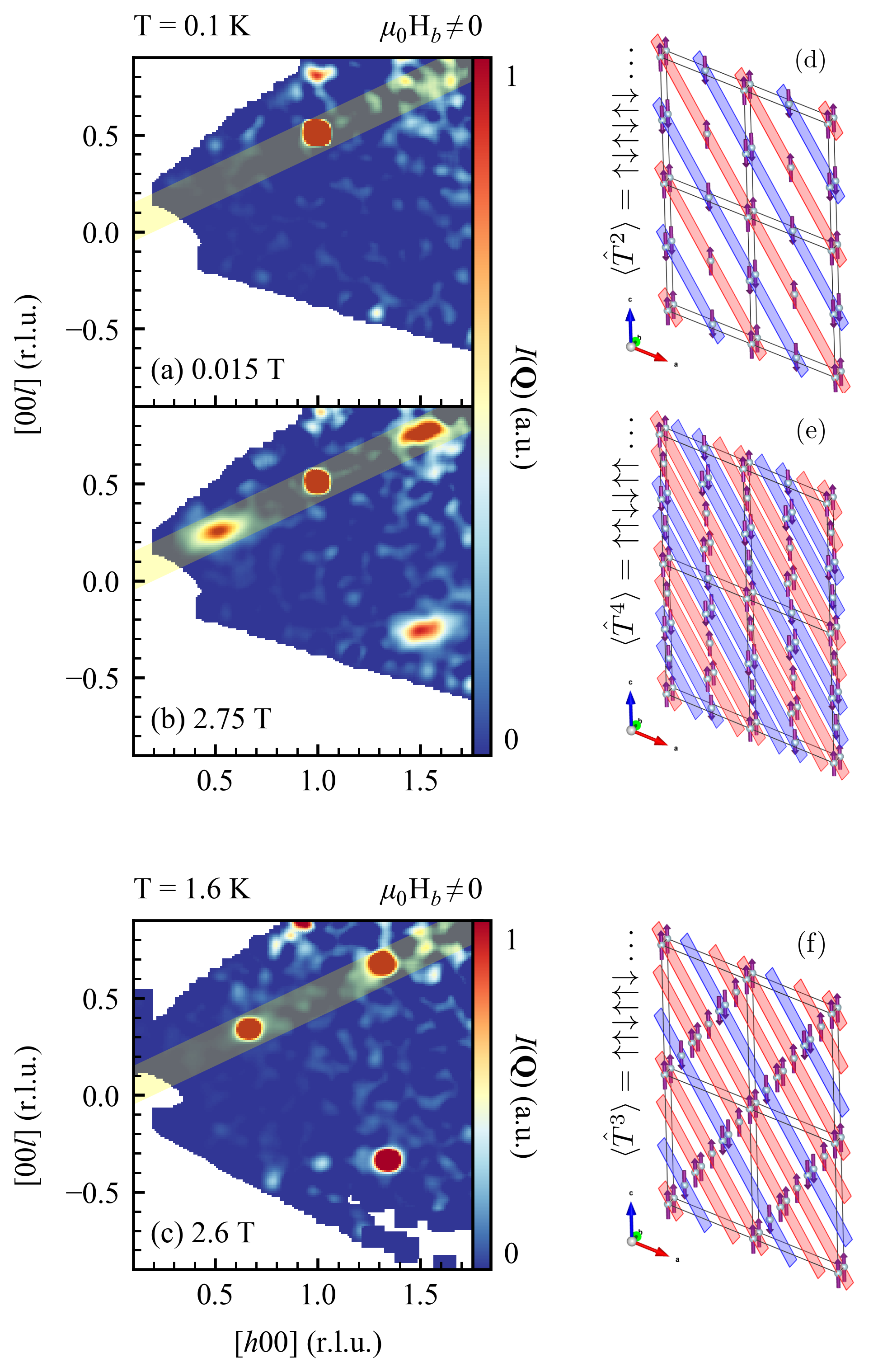}
\caption{\label{fig:fig1}Low temperature (T = 0.1 K) diffraction maps at $(a)$ $\mu_{0}\mathrm{H}=0.015\ \mathrm{T}$ and $(b)$ $\mu_{0}\mathrm{H}=2.75\ \mathrm{T}$ starting from a zero field cooled state. $(c)$ An identical scan at $\mathrm{T}=1.6\ \mathrm{K}$ at $\mu_{0}\mathrm{H}=2.6\ \mathrm{T}$. The highlighted region shows the cut directions. The magnetic periodicities are schematically shown for each scan in panels $(d-f)$, with moment directions fixed to that of the refined zero field structure.}
\end{figure}

Considering the local octahedral field and spin-orbit coupling, the ground state of Co$^{2+}$ can be projected onto a $j_{\mathrm{eff}}={1\over2}$~\cite{Nandi14:118,Sarte18:22} with the ${3\over2}$ spin-orbit levels well separated in energy~\cite{Wallington15:92}.  $\alpha$-CoV$_{2}$O$_{6}$ differs from rocksalt CoO where a large exchange constant induces strong mixing between the $j_{\mathrm{eff}}$ spin-orbit levels~\cite{Cowley13:88,Sarte19:100,Sarte18:98}. The local crystalline distortion further supports an Ising anisotropy in the spin orientation~\cite{Kim12:85,Hollman14:89}, evidenced by a gap in the magnetic dynamics and also the critical scattering discussed below.

We present results of the magnetic field and temperature dependence of the magnetic periodicities in $\alpha$-CoV$_{2}$O$_{6}$ measured with neutron diffraction in single crystals and powders.  This paper is divided into four sections including this introduction.  We first state the experiments and the different instruments used followed by a presentation of the low temperature field dependence of the magnetic periodicities including a comparison at high temperatures.  We then discuss the results on general grounds in the context of the Lieb-Mattis-Shultz (LSM) theorem and compare this analysis to previous theoretical predictions.  

\section{Experiment}

Flux grown single crystals~\cite{He09:131} and powders~\cite{Markkula12:192} were characterized using the MORPHEUS (PSI, Switzerland) and WISH (ISIS, UK) diffractometers. Neutron diffraction experiments under applied magnetic fields used WISH with a vertical magnetic field.  Due to the detector and magnetic field layout on WISH, kinematic constraints requiring access to magnetic Bragg peaks required the field have a significant component along the crystallographic $b$ axis.  From a refinement of the UB matrix from nuclear Bragg positions, the vertical magnetic field had components of 2.1, 15.2, and 85.3$^{\circ}$ along the $a,b,c$ crystallographic axes respectively.  Based on these constraints on WISH, we performed an experiment on the cold triple-axis spectrometer RITA (PSI, Switzerland) using a horizontal magnet where the field could be aligned along the $c$-axis.  Diffraction in a horizontal field and dilution fridge were done with fixed $\mathrm{E}_{i}\equiv \mathrm{E}_{f}=5.0\ \mathrm{meV}$ and with the field aligned along the crystallographic $c$-axis. To reconcile results from the different field geometries, we performed a diffraction experiment on a pressed polycrystalline pellet using TASP (PSI, Switzerland) with $\mathrm{E}_{i} \equiv \mathrm{E}_{f}=3.0\ \mathrm{meV}$. Magnetization were performed using a low temperature SQUID magnetometer at the Institut N\'{e}el in Grenoble~\cite{paulsen_book} and a Physical Properties Measurement System.

\section{Results}

A summary of the WISH results in the $(h0l)$ scattering plane are shown in Fig. \ref{fig:fig1}$(a-c)$ and are compared against real space magnetic periodicities \ref{fig:fig1}$(d-f)$. We note that all data from WISH has had a high magnetic field ($\mu_{0}H$=6 T) scan subtracted where the magnetism is saturated and hence is a measure of the background originating from the sample environment.  Fig. \ref{fig:fig1}$(a)$ plots 0.1 K, 0.015 T data after zero field cooling. The observed $\vec{Q}=(10{1\over 2})$ peak originates from a doubling of the unit cell along the $c$-axis and is drawn in real space showing a $\langle \hat{T}^{2} \rangle =\ \uparrow \downarrow$ periodicity~\cite{Markkula12:86,Lenertz12:86}. On increasing the field to $\mu_{0}\mathrm{H}=2.75$ T (Fig. \ref{fig:fig1}$b$), as well as the $\langle \hat{T}^{2} \rangle =\ \uparrow \downarrow$ peak $\vec{Q}=(10{1\over 2})$, new momentum broadened magnetic peaks at $\vec{k}=({1\over 2} 0 {1\over 4})$ positions are observed, characteristic of spatially short range $\langle \hat{T}^{4} \rangle =\ \uparrow \uparrow \downarrow \downarrow$ or $\uparrow \uparrow \uparrow \downarrow$ translational periodicity. We note that from the momentum dependence alone, there is an ambiguity of the relative arrangement of spins for $\langle \hat{T}^{4} \rangle$ magnetism.  

Fig. \ref{fig:fig1}$(c)$ shows the same high field reciprocal space map but at $\mathrm{T}=1.6\ \mathrm{K}$. In contrast to the 0.1 K data above 2.6 T (Fig. \ref{fig:fig1}$a$), only peaks at $\vec{k}=({2\over 3}0{1\over 3})$ positions are observed characteristic of $\langle \hat{T}^{3} \rangle =\ \uparrow \uparrow \downarrow$ periodicity.  $\langle \hat{T}^{3} \rangle$ is consistent with the results of previously published in-field neutron powder diffraction~\cite{Markkula12:86,Lenertz12:86} and are resolution limited in momentum, characteristic of long range spatial correlations. The spatially short-range $\langle \hat{T}^{4} \rangle$ periodicity is only stabilized in field at low temperatures.  We now present the results of this experiment with first a description of the field dependence at low temperatures where the multiple commensurate periodicities are stablized.  We then compare these results to the high temperature diffraction where only $\langle \hat{T}^{3} \rangle =\ \uparrow \uparrow \downarrow$ periodicity is observed with diffraction measurements.

\subsection{Field Dependence}

\begin{figure}[tb]
\includegraphics[width=75mm]{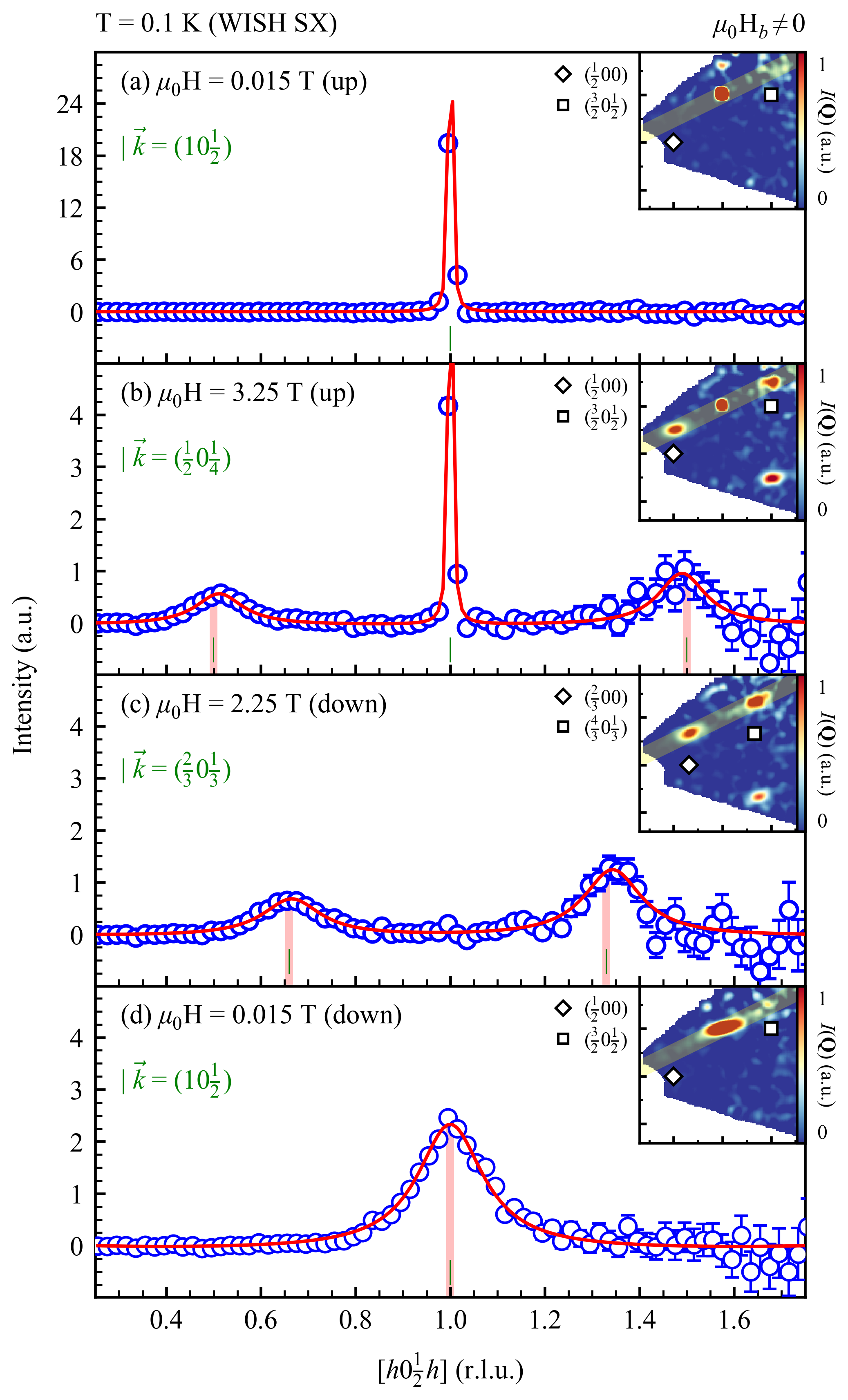}
\caption{\label{fig:fig2} $(\mathrm{T}=0.1\ \mathrm{K})$ scans along $[h0{h\over2}]$ from WISH. On increasing the field momentum broadened $\vec{k}=({1\over2}0{1\over4})$ (panel $b$) peaks draw intensity from $\vec{Q}=(10{1\over 2})$. On decreasing the field from saturation, these are replaced by momentum broadened peaks at $\vec{k}=({2\over3}0{1\over3})$.  At low fields (panel $(d)$), the zero-field structure characterized by $\vec{k}=(10{1\over 2})$, but spatially short ranged.  The solid curves are fits described in the main text, and red bars are the resolution.}
\end{figure}

The different magnetic periodicities from the single crystal WISH data illustrate an uniaxial arrangement of Co$^{2+}$ spins along the highlighted $[h0{h \over 2}]$ direction with $\langle \hat{T}^{2} \rangle$, $\langle \hat{T}^{3} \rangle$, and $\langle \hat{T}^{4} \rangle$ translational symmetry depending on temperature and magnetic field. 
%In Fig. \ref{fig:fig2}, we explore diffraction cuts along this direction in the three magnetic states illustrated in Fig. \ref{fig:fig1} at $\mathrm{T}=0.1\ \mathrm{K}$.
Fig. \ref{fig:fig2} plots diffraction cuts as a function of applied field starting from a zero-field cooled state and increasing the field (termed `up' in Fig. \ref{fig:fig2} $a,b$) through to saturation at 6 T and then on decreasing the field (termed `down' in Fig. \ref{fig:fig2} $c,d$). 

\begin{figure}[tb]
\centering
\includegraphics[width=78mm]{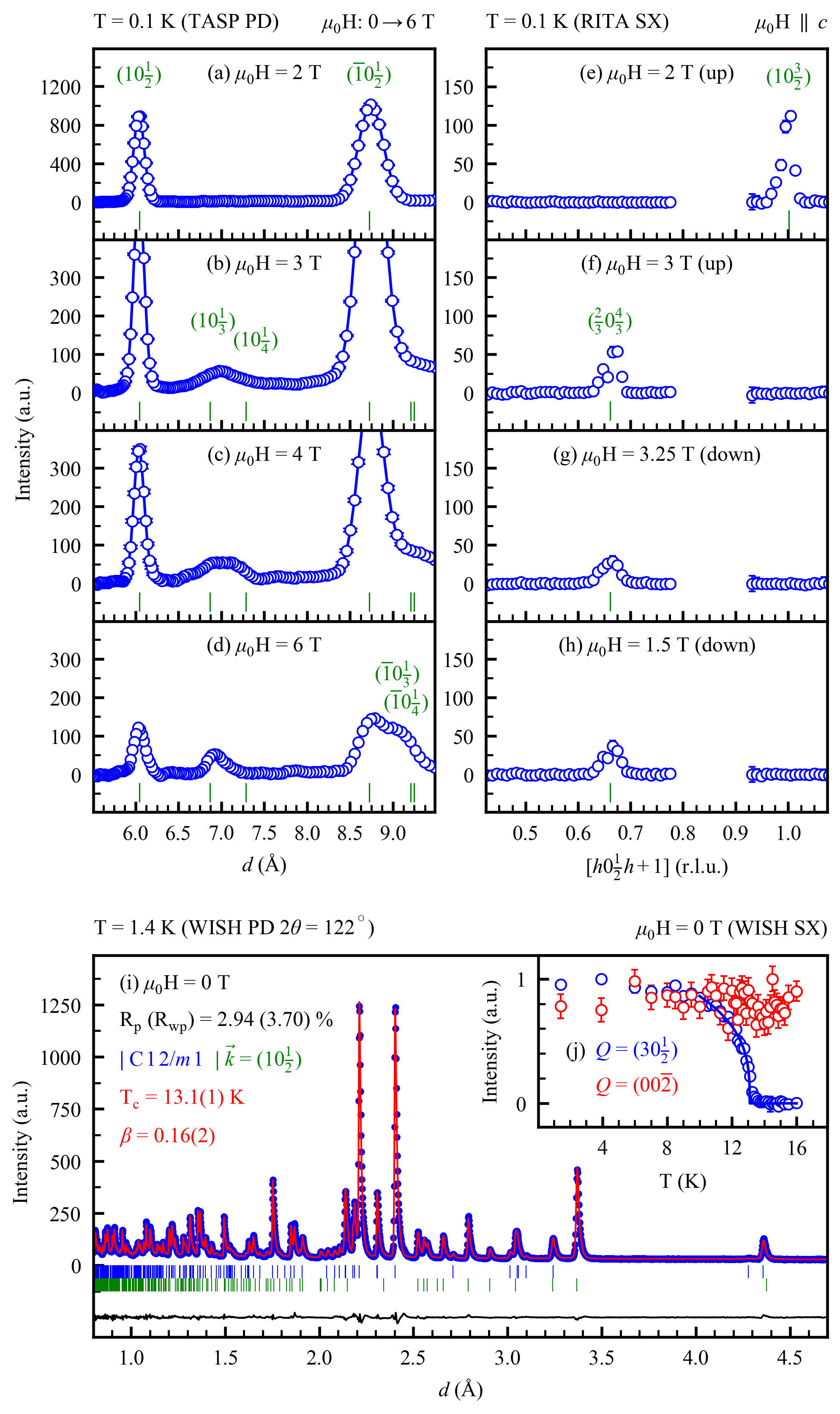}
\caption{\label{fig:fig3}$(a-d)$ powder data from TASP at $\mathrm{T}=0.1\ \mathrm{K}$ with a vertical magnetic field showing the presence of both $\langle \hat{T}^{3} \rangle$ and $\langle \hat{T}^{4} \rangle$. $(e-h)$ single crystal data taken on RITA in the $(h0l)$ plane with the field aligned along the $c$-axis at $\mathrm{T}=0.1\ \mathrm{K}$ showing $\langle \hat{T}^{3} \rangle$ and no observable $\langle \hat{T}^{4} \rangle$ order in this field geometry. $(i)$ shows magnetic and structural characterization of the powder with the inset $(j)$ shows critical scattering described in the text.} 
\end{figure}

\begin{figure*}[tb]
\includegraphics[width=162mm]{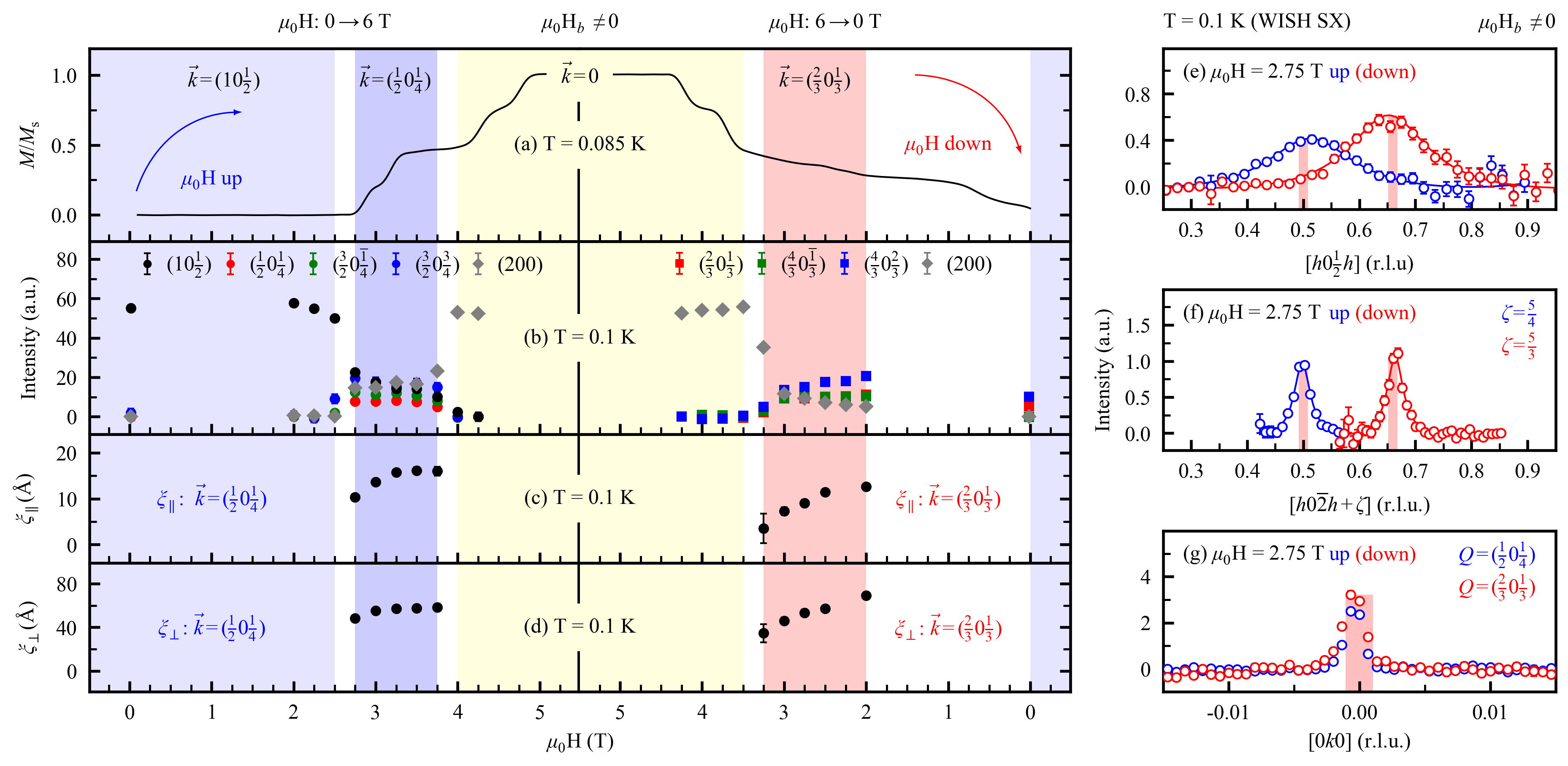}
\caption{\label{fig:fig4}$(a)$ illustrates low temperature magnetization data in the same field arrangement as the WISH diffraction experiments described above. The thermodynamic data is compared against the $(b)$ intensity and $(c,d)$ correlation lengths extracted from WISH single crystal diffraction. Data is presented for field increasing and decreasing experiments where $\langle \hat{T}^{4} \rangle$ and $\langle \hat{T}^{3} \rangle$ periodicities are observed. $(e)$ is a line cut along $[h 0 {h\over 2}]$ (highlighted in Figs. \ref{fig:fig1} $(a-c)$) and compared to cuts perpendicular to this direction both within and perpendicular to the $(h0l)$ scattering plane in panels $(f,g)$.} 
\end{figure*}

The ground state after zero field cooling is characterized by long range magnetic order with commensurate propagation vector ${\vec{k}=(10\frac{1}{2})}$. Upon increasing field $(\mu_0 \mathrm{H} \not\parallel c)$, magnetic intensity is increasingly redistributed from the resolution limited $(10{1\over2})$ $\langle \hat{T}^{2} \rangle$ in Fig. \ref{fig:fig2}$(a)$ among momentum broadened peaks at $({1\over2}0{1\over4})$ and $({3\over2}0{3\over4})$ indicative of spatially short-range $\langle \hat{T}^{4} \rangle$ order. On cycling through to the saturation field and subsequently decreasing the magnetic field Fig. \ref{fig:fig2}$(c)$ illustrates new momentum broadened  $\langle \hat{T}^{3} \rangle$ (at $({2\over3}0{1\over3})$ and $({4\over3}0{2\over3})$) and $\langle \hat{T}^{2} \rangle$ peaks (Fig. \ref{fig:fig2}$(d)$) as the field is decreased, clearly demonstrating a hysteretic degeneracy.

The single crystal results illustrate different $\langle\hat{T}^{q}\rangle$ correlations depending on field orientation and strength.  $\langle \hat{T}^{3} \rangle$ is expected based on previous powder diffraction and reports of ${1\over 3}$ plateaus of the saturated magnetism~\cite{Lenertz11:115,Singh12:22,Xiaoyan12:116,Saul13:87,He14:362,Liu19:31}. We confirm this result by comparison of powder data in Fig. \ref{fig:fig3}$(a-d)$ which shows diffraction data on increasing the field at $\mathrm{T}=0.1\ \mathrm{K}$ from a zero-field cooled state. On increasing the field, we observe a complex series of magnetic peaks consistent with simultaneous $\langle \hat{T}^{2} \rangle$, $\langle \hat{T}^{3} \rangle$, and $\langle \hat{T}^{4} \rangle$ order (green tick marks) due to spherical averaging of the field. We also confirm the Ising nature implied from spectroscopy by considering the powder diffraction (Fig. \ref{fig:fig3} $(i)$) and extract a critical exponent (Fig. \ref{fig:fig3} $(j)$) for $|M|^{2}\propto |T-T_{N}|^{2\beta}$ of $\beta=0.16(2)$.~\cite{Collins:book,Bramwell}

Fig. \ref{fig:fig3}$(e-h)$ illustrates single crystal data (RITA) with the field $||$ to the $c$-axis showing only the stabilization of spatially long-range $\langle \hat{T}^{3} \rangle$ order.  However the WISH single crystal data shows that when the field is rotated away from the $c$-axis, $\langle \hat{T}^{4} \rangle$ periodicity can be stabilized.  We note that $\langle \hat{T}^{4} \rangle=\uparrow \uparrow \downarrow \downarrow$ or $\uparrow \uparrow \uparrow\downarrow$ order can be constructed from un underlying $\langle \hat{T}^{3} \rangle$ order, but with antiphase boundaries~\cite{Stock17:119}. This can be seen by considering the example $\langle \hat{T}^{3+} \rangle = \uparrow \uparrow \downarrow$ and $\langle \hat{T}^{3-} \rangle = \downarrow \downarrow \uparrow$.  $\langle \hat{T}^{3+} \hat{T}^{3+} \hat{T}^{3-} \hat{T}^{3-}\rangle= \uparrow \uparrow \downarrow \uparrow \uparrow \downarrow \downarrow \downarrow \uparrow \downarrow \downarrow \uparrow \equiv \langle  \hat{T}^{3+}  \hat{T}^{4}  \hat{T}^{2}  \hat{T}^{3-}\rangle$.  Therefore coexisting $\langle \hat{T}^{2,3,4} \rangle$ can be created with  $\langle \hat{T}^{3} \rangle$ order with antiphase boundaries.  Given the $j_{eff}$=${1 \over 2}$ Ising nature of the magnetic moments, we note that the antiphase boundaries in $\alpha$-CoV$_{2}$O$_{6}$ are expected to be spatially sharp in contrast to boundaries which are suggested to exist in large spin magnets such as CaFe$_{2}$O$_{4}$~\cite{Haldane83:50,Stock16:117,Stock16:117,Lane20:102} which are believed to be extended in space.   We note that due to the insulating nature of $\alpha$-CoV$_{2}$O$_{6}$, we interpret these new periodicities in terms of local moment conserving antiphase boundaries and not in terms of a density wave as observed in itinerant materials.~\cite{Rodriguez11:83,Stock17:95}   This situation has been discussed recently in the context of the magnetic order in Cu$_{3}$Nb$_{2}$O$_{8}$.~\cite{Giles20:102}   $\langle \hat{T}^{4} \rangle$ order is stabilized when the field is rotated away from the $c$-axis allowing local antiphase boundaries to form and $\langle \hat{T}^{4} \rangle$ periodicity to emerge.  

\begin{figure*}[tb]
\includegraphics[width=162mm]{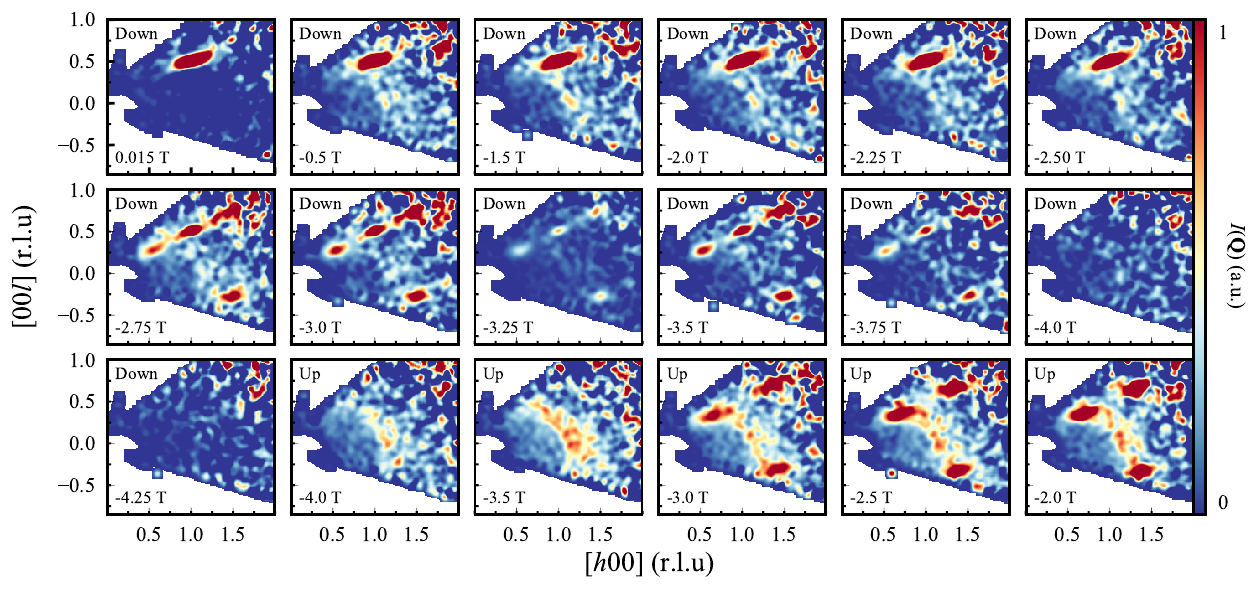}
\caption{\label{fig:fig6} T=0.1 K WISH (ISIS) data in the $(h0l)$ plane for negative fields taken after a  $\mu_{0}H \rightarrow$ 6 T $\rightarrow$ 0 T field cycle.  On decreasing field commensurate $\langle \hat{T}^{2} \rangle = \uparrow\downarrow$ and $\langle \hat{T}^{4} \rangle$ spatially localized correlations exist.  On increasing the field back to 0 T short-range $\langle \hat{T}^{3} \rangle$  is recovered.  The results further illustrate strong hysteretic effects at low temperatures in $\alpha$-CoV$_{2}$O$_{6}$. }
\end{figure*}

The  multiple, metastable, commensurate periodicities are summarized  in Fig. \ref{fig:fig4} from WISH and compared to low-temperature ($\mathrm{T}=0.085\ \mathrm{K}$) magnetization (Fig. \ref{fig:fig4} $a$) in a similar field geometry,  sensitive to the Q=0 response. The field is cycled from $\mu_{0}\mathrm{H}=0\ \mathrm{T}$ up to saturation of the magnetization and then decreased, illustrated by the two halves. In Fig. \ref{fig:fig4}$(b-d)$, the intensity and correlation lengths are plotted based on a fit to lattice-Lorentzians~\cite{Guinier:book,Zal:book} characteristic of exponentially decaying spatial correlations parallel ($\xi_{\parallel}$) and perpendicular ($\xi_{\perp}$) to the scan direction shown in Fig. \ref{fig:fig1},

\begin{equation}
I(\vec{q})\propto{{\sinh(a_{\alpha}\xi_{\alpha}^{-1})} \over {\cosh(a_{\alpha}\xi_{\alpha}^{-1})-\cos(\vec{q}\cdot \vec{a}_{\alpha})}}. \nonumber
\end{equation}

\noindent where the correlation length along a particular direction $\vec{a}_{\alpha}$ is given by $\xi_{\alpha}$.    We note that for the diffraction experiments on WISH, the experiments are energy integrating therefore providing a measure of $\int dE S(Q,E)\equiv S(Q)$, therefore the correlation lengths are the instantaneous values.~\cite{Collins:book}

\begin{figure*}[hbtp]
\includegraphics[width=146mm]{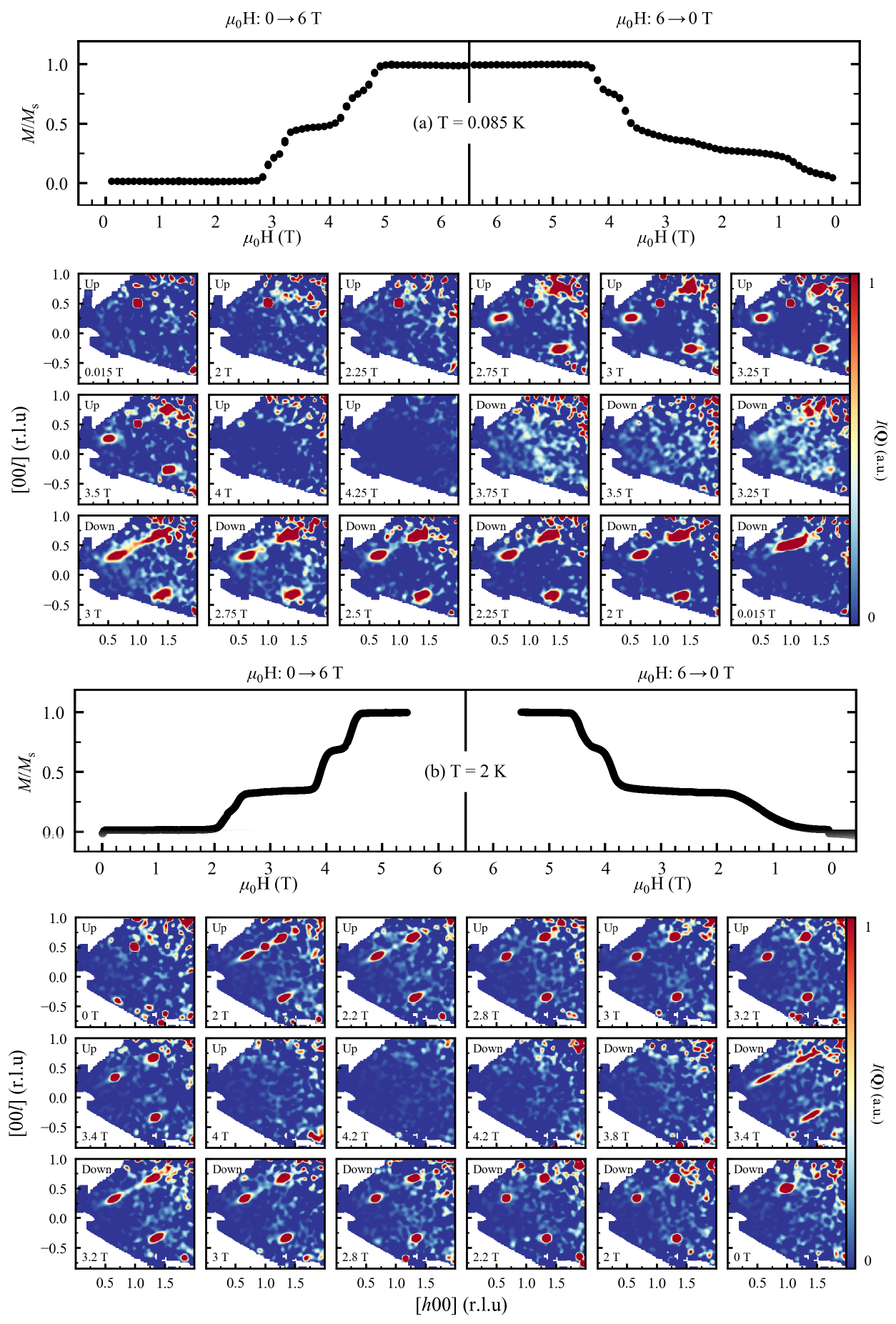}
\caption{\label{fig:fig1_sup}Field dependence of the bulk magnetization and diffuse neutron scattering in the $(h0l)$ plane for (a) T $\leq$ 100 mK and (b) T $\leq$ 2 K. Below 100 mK, one finds a shift of the magnetic propagation vector from $\vec{k} = (\frac{1}{2}\,0\,\frac{1}{4})$ to $\vec{k} = (\frac{2}{3}\,0\,\frac{1}{3})$ upon cycling the field through $\pm \mu_0$H values, while at higher temperatures the propagation vector $\vec{k} = (\frac{2}{3}\,0\,\frac{1}{3})$ is unchanged.}
\end{figure*}

Fits are shown in Figs. \ref{fig:fig4} $(e)$ along the $[h0{h \over 2}]$ direction highlighted in Fig. \ref{fig:fig1} and perpendicular to (panels $f,g$) with the resolution width defined by the vertical bars. $\langle \hat{T}^{4} \rangle$ periodicity is observed on increasing the field with spatially anisotropic correlations. Long-range spatial correlations are observed along the $[0k0]$ direction while very short range correlations of just a few unit cells are seen along the stacking direction (Fig. \ref{fig:fig1}).  $\langle \hat{T}^{4} \rangle$ order is only observed to be stabilized on increasing the magnetic field from a zero-field cooled state. On decreasing the field from magnetic saturation, short-range $\langle \hat{T}^{3} \rangle$ order is observed with a similar spatial anisotropy to the $\langle \hat{T}^{4} \rangle$ order. We emphasize that this response is only observed at low temperatures, at $\mathrm{T}=1.6\ \mathrm{K}$ (Fig. \ref{fig:fig1}) only spatially long range $\langle \hat{T}^{3} \rangle$ magnetic order is observed and is further discussed in the next section.

Fig. \ref{fig:fig4} demonstrates a hysteresis at low temperatures when cycling the field through saturation with $\langle \hat{T}^{4} \rangle$ stabilized on increasing the field but $\langle \hat{T}^{3} \rangle$ and finally $\langle \hat{T}^{2} \rangle$ on decreasing.  This is further shown in Fig. \ref{fig:fig6} which illustrates T=0.1 K diffraction data in the $(h0l)$ plane on WISH after cycling the field through $\mu_{0}H\rightarrow$ 6 T $\rightarrow$ 0 T at negative fields. On decreasing the field to $\mu_{0}H$=-6 T, there is a region where $\langle \hat{T}^{4} \rangle$ and $\langle \hat{T}^{2} \rangle$ spatial local correlations exists.  On increasing the field from $\mu_{0}H$=-6 T $\rightarrow$ 0 T, spatially short-range $\langle \hat{T}^{3} \rangle$ are observable on continuing to increase the field to 0 T.  These results are indicative of a strong hysteretic nature to the magnetic correlations at low temperatures with the magnetic spatial correlation length is only a few unit cells along the stacking direction and coexists with other periodicities.  While our results are at finite temperature, we observe multiple instabilities on changing magnetic field at low temperatures far below the onset of N\'{e}el order (Fig. \ref{fig:fig3}).  We emphasize that the spatially short range order found here cannot be interpreted as magnetic field induced phase transitions as is the case in the examples mentioned in the introduction to this paper due to the absence of a resolution limited Bragg peak associated with the establishment of long-range order.

\subsection{Temperature dependence of the diffuse scattering}

We now compare the low temperature field dependence of the neutron diffraction and magnetization to high temperatures where only $\langle \hat{T}^{3} \rangle$ periodicity is observed (Fig. \ref{fig:fig1}).  The comparison is illustrated in Fig. \ref{fig:fig1_sup}.  For $\mathrm{T} \leq 100\ \mathrm{mK}$, increasing applied magnetic field from 0 T results in the stabilization of a short-range ordered state with $\langle \hat{T}^{4} \rangle$ translational symmetry. The magnetic diffuse scattering in this state is anisotropic, with enhanced broadening of the component along the antiferromagnetic propagation vector direction $[1\,0\,\frac{1}{2}]$ compared to the perpendicular component along $[h\,0\,2\overline{h}]$. This implies enhanced disorder along the antiferromagnetic propagation vector direction and supports the picture of stacking faults between ferromagnetic planes along this direction as discussed above.  After saturation at 6 T in the collinear ferromagnetic state, decreasing field results in a shift of the propagation vector from $\vec{k} = (\frac{1}{2}\,0\,\frac{1}{4})$ to $\vec{k} = (\frac{2}{3}\,0\,\frac{1}{3})$ and the stabilization of a state with $\langle \hat{T}^{3} \rangle$ translational symmetry. Magnetic intensity is increasingly redistributed along the antiferromagnetic propagation vector direction implying reduced correlation length on the return leg of the field sweep. Returning to 0 T, the $\vec{k} = (1\,0\,\frac{1}{2})$ state has developed a reduced correlation length on account of the broad $Q = (1\,0\,\frac{1}{2})$ reflection. Conversely, at ${\mathrm{T} = 2\ \mathrm{K}}$, no shifting of the propagation vector on cycling field through $\pm \mu_0$H is observed, with the $\langle \hat{T}^{3} \rangle$ state manifest on both up and down field sweeps and comparatively only very minor broadening of the $Q = (1\,0\,\frac{1}{2})$ reflection.

\section{Discussion}

Theoretical work performed on the Ising model~\cite{Fisher80:44} report the existence of an infinite number of commensurate, and collinear, magnetic phases at low temperatures based on the presence of antiphase boundaries. This was found to be supported when the anisotropic next nearest neighbor interaction (ANNNI) satisfies ${-J_{2} \over J_{1}}>{1\over 2}$. With decreasing temperature, we observe that both $\langle \hat{T}^{3} \rangle$ ($\uparrow \uparrow \downarrow$) and spatially short range $\langle \hat{T}^{4} \rangle$ ($\uparrow \uparrow \downarrow \downarrow$) can be stabilized. It will be interesting for lower temperatures to be pursued to determine if further periodicities can be be stabilized. Indeed the low temperature magnetization data shown in Fig. \ref{fig:fig4}$(a)$ is suggestive of higher order plateaus at fractional values of the saturated magnetization. Judging from the spatially short-range nature of the $\langle \hat{T}^{4} \rangle$ periodicity with neutron diffraction, such phases may be too weakly correlated or short-lived to observe~\cite{Kudasov18:185}.  In this context of microscopic exhange and anisotropy energies, these results may suggest that strongly Ising-like $\alpha$-CoV$_{2}$O$_{6}$ may have the correct balance of exchange couplings to support an infinite series of commensurate phases with decreasing temperature. 

However, $\alpha$-CoV$_{2}$O$_{6}$ displays low temperature magnetization plateaus built upon metastable, spatially short-range, and commensurate magnetic correlations that do not break long-range translational symmetry owing to the finite spatial correlation lengths.  These are not phase transitions with a well defined Bragg peak in these plateaus and it is difficult to understand these results in terms of the competing exchange interactions as reported in some localized magnets (for example in Ref. \onlinecite{Fishman10:81, Chen16:28}) and discussed in the previous paragraph. The presence of a plateau is indicative of an energetic gap, yet translational symmetry is not broken in $\alpha$-Co$_{2}$V$_{2}$O$_{6}$.  We now discuss this apparent dichotomy. 

Following Ref. \onlinecite{Oshikawa97:78}, on general grounds, without specifying the particular form of the microscopic magnetic Hamiltonian, the origin of discrete steps in the magnetization can be understood  from the underlying translational symmetry of the Hamiltonian. This underlying symmetry places restrictions on the values of the complex geometric phase acquired when rotating the spins around an axis of symmetry. For a Hamiltonian invariant under a site translation $\hat{\mathcal{H}}\mapsto\hat{T}\hat{\mathcal{H}}\hat{T}^{-1}$, a norm preserving unitary rotation of the spins generates a geometric phase under the gauge transformation

\begin{equation}
\hat{T}\hat{U}\hat{T}^{-1}=\hat{U}e^{2\pi i \left(S-m\right)},
\label{eqn:phase}
\end{equation}

\noindent where $m={1\over L} \sum^{L}_{i=1}S_{i}^{z}$. This is akin to that gained by a Foucault pendulum as it undergoes a unitary transformation about the Earth's axis. For $(S-m) \notin \mathbb{Z}$, the rotated state must be orthogonal to the ground state and represent an excited state with an infinitesimally larger energy in the thermodynamic limit~\cite{Oshikawa97:78}. These states lead to a gapless continuum as Haldane suggested~\cite{Haldane83:93A}. In the case of integer values, states are gapped and cannot be smoothly transformed through with the application of field, leading to the emergence of magnetization plateaus and a Bloch state analogue of the quantum Hall effect.~\cite{Stormer99:71,Laughlin81:23,Oshikawa97:78}  This is the Lieb-Mattis-Shultz theorem\cite{Lieb61:16} which states that a spin chain can have an energy gap without breaking translational symmetry when the magnetization per spin satisfies $(S-m)\in \mathbb{Z}$. In the case that $(S-m)= p/q$, where $p$ and $q$ are coprime, $\hat{T}^{q}$ preserves gauge invariance and the ground state is endowed with $\hat{T}^{q}$ symmetry. This new magnetization-induced symmetry can give rise to modulated ground states, though the stability of such states is ultimately determined by the interaction terms present in the microscopic Hamiltonian. 

The presence of a strong spin-orbit splitting between the $j_{\mathrm{eff}}={1\over 2}$ ground state doublet and the excited $j_{\mathrm{eff}}={3\over 2}$ and $j_{\mathrm{eff}}={5\over 2}$ multiplets allows the inter-site interaction Hamiltonian to be written in terms of effective spin-half operators, $\mathbf{\tilde{S}}_{i}=\gamma \mathbf{\tilde{j}}_{i}$, where $\gamma$ is a projection factor.~\cite{Abragam:book} Using the general arguments discussed previously, one can assign a value of $p/q$ to each of the magnetization plateaus and hence extract the translational symmetry of the ground state.

\begin{table}[h]
\caption{\label{plateaus} Values of $(S-m)=p/q$ for each of the magnetization plateaus observed in Fig. \ref{fig:fig4} (a).}
\begin{ruledtabular}
\begin{tabular}{ccc}
$S-m =p/q$ & $M/M_{S}$ & $\langle \hat{T}^{q} \rangle$ \\ 
\hline
$\frac{1}{2} - \frac{1}{2} =0 $ & 1 & $\langle \hat{T}^{1} \rangle =\uparrow \uparrow$  \\ 
$\frac{1}{2} - 0 =\frac{1}{2} $ & 0 & $\langle \hat{T}^{2} \rangle =\uparrow \downarrow$ \\ 
$\frac{1}{2} - \frac{1}{6} =\frac{1}{3} $ & $\frac{1}{3}$ & $\langle \hat{T}^{3} \rangle =\uparrow \uparrow \downarrow$ \\ 
$\frac{1}{2} - \frac{1}{4} =\frac{1}{4} $ & $\frac{1}{2}$ & $\langle \hat{T}^{4} \rangle =\uparrow \uparrow \downarrow \downarrow, \uparrow \uparrow \uparrow \downarrow$   \\[1ex] 
%\hline
\end{tabular}
\end{ruledtabular}
\end{table}

Regarding the hysteresis reported in Figs. \ref{fig:fig4} and \ref{fig:fig6}, given the insulating nature~\cite{Oshikawa03:90} and the requirement of gauge invariance, this change in the ground state, despite completing a closed loop of unitary transformations, maybe indicative of a non-trivial underlying topological structure.  We do not expect a violation of the adiabatic theorem given the large excitation gap ($\sim$ 1 meV) in comparison to the applied fields (1 T $\sim$ 0.1 meV), however a highly degenerate ground state is expected for Ising spins on a triangular motif~\cite{Stephenson64:5}. In further analogy to the Hall effect, we note that hysteresis effects have been reported in the fractional quantum Hall effect~\cite{Cho98:81,Ihna07:75}.  The hysteresis here is distinct from the very slow critical behavior observed in Ca$_{3}$Co$_{2}$O$_{6}$~\cite{Agrestini11:106} where a phase transition occurs, yet taking at least hours.

In summary, we have applied neutron diffraction and magnetization in an applied magnetic field to report the presence of multiple commensurate periodicities in an Ising $j_{\mathrm{eff}}$=${1\over 2}$ magnet. We find spatially short range and metastable commensurate magnetic periodicities in the magnetization plateaus. We have understood these spatially localized periodicities in terms of the Lieb-Mattis-Shultz thereom.  $\alpha$-CoV$_{2}$O$_{6}$ represents an example of Ising magnet displaying multiple commensurate and metastable local periodicities defining plateaus in the magnetization.

\begin{acknowledgements}

We are grateful for funding from the EPSRC, STFC, the Royal Society, and the Carnegie Trust for the Universities of Scotland.

\end{acknowledgements}

%\bibliography{CVO_bibliography}


\end{document}